\documentclass[
 aip,
 amsmath,amssymb,
 reprint,%
groupedaddress
]{revtex4-1}

\usepackage{graphicx}
\usepackage{dcolumn}
\usepackage{bm}
\usepackage{hyperref}

\usepackage[utf8]{inputenc}
\usepackage[T1]{fontenc}
\usepackage{mathptmx}
\usepackage{etoolbox}
\usepackage{xcolor}

\newcommand{\Google}{\affiliation{Google Quantum AI, Goleta CA 93117, USA}}
\newcommand{\Keysight}{\affiliation{Keysight Technologies, 1400 Fountaingrove Pkwy., Santa Rosa, CA 95403, USA}}

\begin{document}
\begin{title}


\title{Modeling flux-quantizing Josephson junction circuits in Keysight ADS}

\author{Ofer Naaman}\email{ofernaaman@google.com}\Google
\author{Mohamed Awida Hassan}\email{mohamed.hassan@keysight.com} \Keysight
\author{Ted White}\Google
\author{Derek Slater}\Keysight
\author{Sean Mcilvane}\Keysight
\author{Edwin Yeung}\Keysight
\author{Philip Krantz}\Keysight
\date{\today}

\begin{abstract}
We introduce Josephson junction and inductor models in Keysight ADS that feature an auxiliary flux port, and facilitate the expression of flux quantization conditions in simulation of superconducting microwave circuits. We present several examples that illustrate our methodology for constructing flux-quantizing circuits, including dc- and rf-SQUIDs, tunable couplers, and parametric amplifiers using SNAIL and rf-SQUID arrays. We perform DC, S-parameter, and harmonic balance simulations to validate our models and methods against theory and published experimental results.
\end{abstract}

\maketitle
\end{title}

\section{Introduction}
In superconducting circuits, the fundamental dynamical variable is the magnetic flux $\Phi=\int{V(t)dt}$, i.e.~the time integral of the voltage, rather than the voltage itself. Circuit simulators in most standard Electronic Design Automation (EDA) tools solve for node voltages rather than fluxes, dropping in the process the integration constant, which constrains inductive superconducting loops to enclose an integer number of flux quanta, $\Phi_0=h/2e=2.068\times 10^{-15}$~Wb (equivalent to $2.068\;\mathrm{mV}\cdot\mathrm{ps}$ or, alternatively, $2.068\;\mathrm{mA}\cdot\mathrm{pH}$), where $h$ is Planck's constant and $e$ is the electron charge. This constraint on the circuit solution, referred to as `flux quantization', is not respected in standard EDA tools, unless the circuit is time-evolved  to its operating point in transient mode from a consistent initial condition (for example, all voltages, currents, and sources set to zero at $t=0$). The absence of flux quantization in standard EDA circuit simulators is one of the main challenges to superconducting circuit designers using these tools.

Superconducting technology additionally introduces a new circuit element, the Josephson junction. It is an indispensable element within the realm of superconducting electronics\cite{barone1982physics, van1999principles}, serving as enabling foundation for a myriad of applications spanning quantum computing\cite{krantz2019apr}, high-speed electronics\cite{Likharev1991IEEE}, metrology\cite{Benz2004IEEE}, and sensing\cite{Qiu2023natphys}. The absence of the Josephson junction from most EDA tools' standard component libraries is another challenge for superconducting circuits designers working within these application spaces.

Designers of digital superconducting circuits\cite{mukhanov2011energy,herr2011ultra,takeuchi2022adiabatic} adopted specialized SPICE variants, such as WRSpice\cite{whiteley1991josephson}, that support accurate transient simulations with Josephson junctions models. More recently, these tools have also gained adoption by designers of superconducting parametric amplifiers\cite{naaman2019high, malnou2024traveling, ranadive2024traveling}, despite the inconvenience of using transient simulations for tasks that are better suited for S-parameters or harmonic-balance (HB) simulations. Other specialized tools for HB simulations of Josephson junction circuits and amplifiers\cite{obrien2022josephsoncircuits, Peng2022IEEE} have been recently introduced, as well as Josephson junction models for Keysight Advanced Design System (ADS)\cite{choi2023IEEE,shiri2023modeling}. Other designers have used various approximations and closed-form models to perform S-parameters and HB simulations of Josephson parametric amplifiers using standard component libraries in ADS\cite{kaufman2023josephson}. All these illustrate the need for a modern and user-friendly circuit simulation environment that includes useful Josephson junction models, is aware of flux quantization constraints, and is capable of performing simulations in the range of modalities relevant to microwave superconducting and Josephson junction circuits.

Starting with release 2022-U1, Keysight ADS has added Josephson junction models in their standard component library. In what follows, we will describe these models, as well as additional models and methods that facilitate accurate simulations of flux-quantizing superconducting circuits. We will give several examples to demonstrate the use of these components, going from simple circuits like the rf- and dc-SQUID\cite{barone1982physics} to more complicated circuits like the `snake'\cite{white2022readout} and SNAIL\cite{frattini20173} nonlinear elements, and impedance matched parametric amplifiers\cite{kaufman2023josephson}.

\section{Josephson Junction models in ADS}

\begin{figure}
    \centering
    \includegraphics[width=\columnwidth]{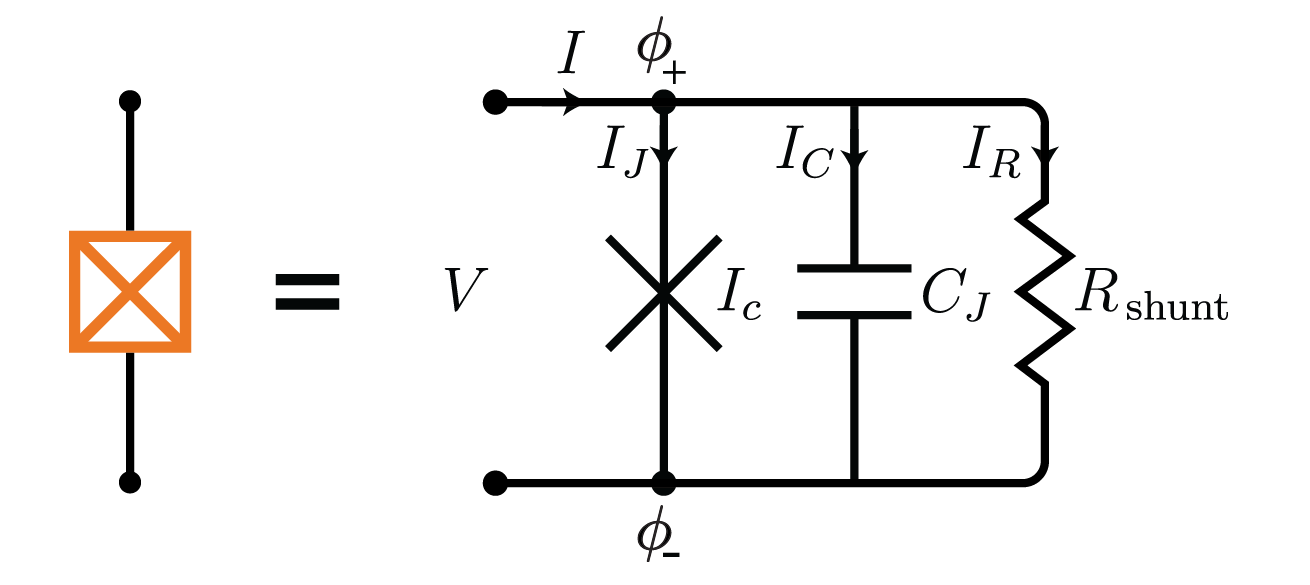}
    \caption{The RCSJ model for Josephson junctions. The node fluxes are normalized to the flux quantum: $\phi_\pm=\Phi_\pm/\Phi_0$.}
    \label{fig:RCSJ}
\end{figure}

 The behavior of Josephson junctions can be accurately described by the Resistively and Capacitively Shunted Josephson junction (RCSJ) model\cite{barone1982physics}, shown in Fig.~\ref{fig:RCSJ}, which incorporates the dynamics of the superconducting phase difference across the junction. In this model, the junction's behavior is governed by a set of nonlinear differential equations representing the conservation of energy and charge, enabling the prediction and analysis of its electrical properties under different operating conditions. The RCSJ model serves as a fundamental tool in understanding and designing Josephson junction-based circuits.

The supercurrent flowing through the junction is related to the Josephson phase, $2\pi\phi$, by the current-phase relation:
\begin{equation} \label{eq:JJ_current}
I_J = I_c \sin{(2\pi\phi)},
\end{equation}
where $I_c$ is the junction's critical current. The Josephson phase, or the superconducting phase-difference across the junction, is essentially an angular measure of the flux $\Phi$ along the junction branch, normalized so $2\pi\phi=2\pi\Phi/\Phi_0$. As we will see below, it is more convenient here to work with node-fluxes instead of branch fluxes; the reduced branch flux, $\phi=\phi_+-\phi_-$, is the difference of normalized node fluxes across the junction, $\phi_\pm=\Phi_\pm/\Phi_0$ in Fig.~\ref{fig:RCSJ}. The voltage across the junction, $V$ in Fig.~\ref{fig:RCSJ}, is related to the time derivative of the reduced flux $\phi$ by the AC Josephson relation:
\begin{equation} \label{eq:JJ_voltage}
V = \Phi_0\frac{d\phi}{dt}.
\end{equation}

The RCSJ model combines the current flowing in the junction Eq.~(\ref{eq:JJ_current}) with those flowing in the resistance and capacitance, resulting in the following equation for the total current $I=I_J+I_R+I_C$:
\begin{equation} \label{eq:JJ_current_RCSJ}
I = I_c \sin{\left(2\pi\phi\right)} + \frac{V}{R_{\mathrm{shunt}}} + C_J \frac{dV}{dt}.
\end{equation}
Substituting the AC Josephson relation in Eq.~(\ref{eq:JJ_voltage}) into Eq.~(\ref{eq:JJ_current_RCSJ}), yields the usual second-order differential equation\cite{barone1982physics} for $\phi$, however, the pair of first-order equations above is better suited for use in circuit simulators. 

To capture this behavior in circuit simulation, ADS uses an auxiliary port on the junction schematic, and expresses the normalized node-flux difference across the junction, $\phi$, as a fictitious voltage $V_\phi=\phi$ appearing on this port. Directly relating the phase difference across the junction to a voltage, such that $1\,$V is equivalent to a $2\pi$ phase difference, allows us to repurpose the simulator's native solvers to account for fluxes in a `flux circuit' just as it would account for voltages in an electrical circuit. The Josephson junction components (and the superconducting inductor components in Sec.~\ref{sec:inductors}) serve to bridge between, and translate from, the electrical circuit to the flux circuit via Eq.~(\ref{eq:JJ_voltage}). With this, the system can be represented solely through equations encompassing currents and voltages at distinct nodes.
Equation~(\ref{eq:JJ_current_RCSJ}) can then be used to capture the dynamics of the Josephson junction under the influence of external bias currents and voltages, resistance, and capacitance. 

We will see below that exposing the node $V_\phi$ (either as a ground-referenced single-ended or as a floating differential voltage) on the `phi' terminal of the junction models in ADS, allows the designer to enforce a phase- or flux-bias on the circuit, and opens the way to account for flux quantization conditions where appropriate in circuits containing junctions and inductors. Note that by convention, current flowing into the junction's positive terminal corresponds to a positive phase-difference, and hence a positive $V_\phi$.

It is worth noting here that Josephson junctions could be produced with a shunt resistance in order to control hysteresis effects. In Keysight ADS, when the model parameters are set to $R_\mathrm{shunt}$=0 and $\beta_C$=0, it implies the absence of shunt resistance by default, resulting in a hysteretic device. The ADS models allow the specification of a shunt resistor either directly through the $R_{\mathrm{shunt}}$ parameter, or indirectly through a specification of the McCumber damping factor $\beta_C$ such that $R_{\mathrm{shunt}}$ is calculated as 
\begin{equation}
R_{\mathrm{shunt}} =
\begin{cases}
R_{\mathrm{shunt}} & \text{if } R_{\mathrm{shunt}} > 0 \\
\sqrt{\frac{\beta_C \Phi_0}{2\pi I_c C_J}} & \text{if } R_{\mathrm{shunt}} = 0 \text{ and } \beta_C>0.
\end{cases}
\end{equation}
In many cases, for unshunted junctions it is useful to set $R_\mathrm{shunt}$ to some high, but finite resistance (e.g.~$1\,\mathrm{M}\Omega$), to improve convergence of the simulation. 

\begin{figure}[t]
    \includegraphics[width=\columnwidth]{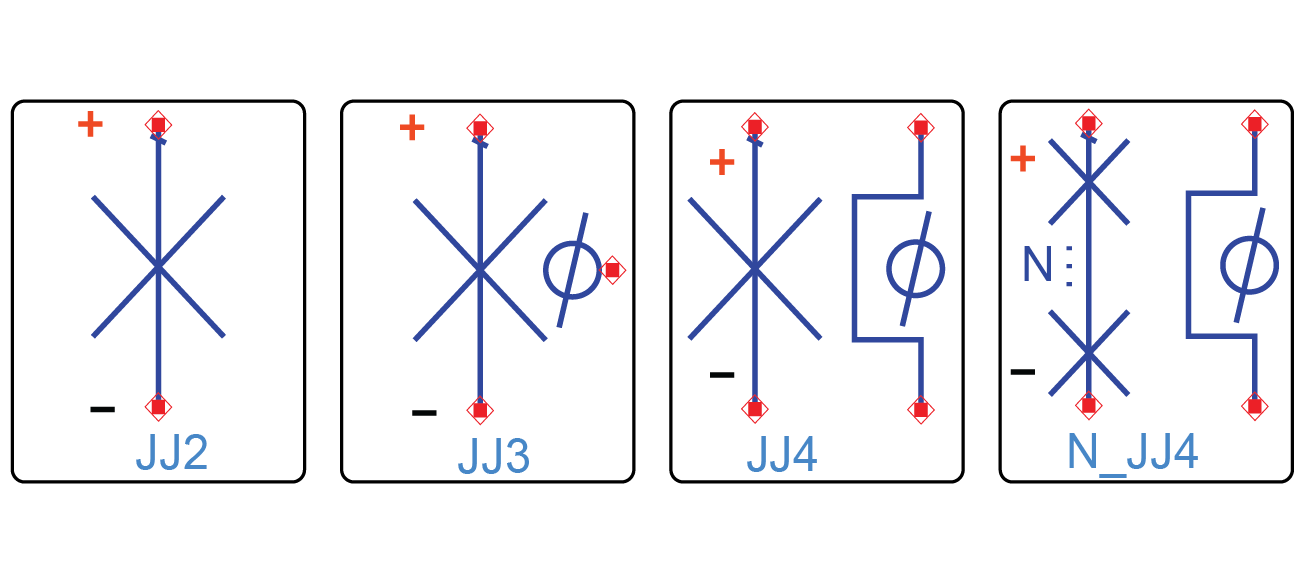}
    \caption{Josephson junction models in ADS 2025. See text for descriptions.\label{fig:jj_models}}
\end{figure}

There are two basic behavioral models of the Josephson junctions in the Keysight ADS:
\begin{itemize}
    \item Two-terminal device (JJ2): follows the above RCSJ model while leaving the $V_\phi$ node floating and not exposed. In this case, the junction is driven by its supercurrent, and the phase adjusts according to Eq.~(\ref{eq:JJ_current}).
    \item Three-terminal device (JJ3): again follows the above RCSJ model but exposes the $V_\phi$ node. In this case, the junction can be driven from the $V_\phi$ node, effectively mimicking a flux biased junction. For the three-terminal device, the single-ended voltage at the $V_\phi$ node is ground-referenced.
\end{itemize}

To support the use of the flux nodes to implement the flux quantization condition into the simulated circuit, the following Josephson Junction models were added in the ADS 2025 release, 
\begin{itemize}
    \item Four-terminal device (JJ4): here the flux has two terminals and the voltage drop between them results in \(V_\phi\) that in turn controls the phase drop across the junction.
    \item An array of \(N\) identical junctions (N-JJ4): The voltage drop across the \(N\) junctions is scaled by their number \(N\) such that
    \begin{equation}
    V_\phi = N V_{\phi\text{-JJ}},
    \end{equation}
    where $V_{\phi\text{-JJ}}$ is the voltage drop across the phi nodes of the individual Josephson junctions.
\end{itemize}
The various junction models and their schematic symbols are depicted in Figure~\ref{fig:jj_models}. Note that \texttt{JJ3} is equivalent to a \texttt{JJ4} with its negative phase terminal connected to ground. Alternatively, \texttt{JJ4} can be constructed from \texttt{JJ3} by balancing its phi port using an ideal $1:1$ transformer.

\section{Inductance Models with Flux Nodes}\label{sec:inductors}
Inductance plays a crucial role in inducing flux within superconducting loops, and this flux has to be accounted for if we want to accurately simulate flux quantizing circuits. A current $I$ flowing through an inductor $L$ will induce a flux $\Phi=LI$ across the inductor's terminals. This can be accounted for by adding a ``flux'' port to the standard inductor model, and outputting a voltage $V_{\Phi}=LI/\Phi_0$ volts across this second port.

\begin{figure}[h]
    \includegraphics[width=\columnwidth]{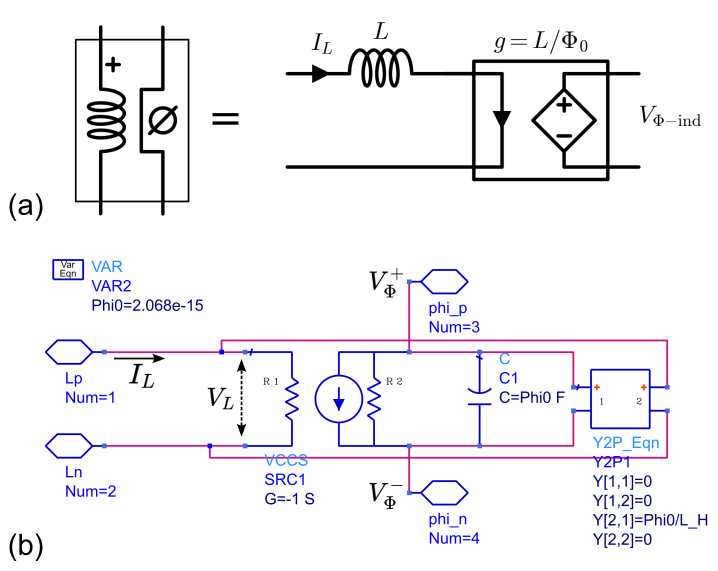}
    \caption{(a) Basic flux-aware inductor model equivalent circuit using a current-controlled voltage source, $V_{\Phi\text{-ind}}=gI_L$, where the transimpedance is $g=L/\Phi_0$ Ohms. (b) A comprehensive linear inductor model following the methods of Ref.~\onlinecite{shiri2023modeling}. \label{fig:L_phase_equiv}}
    
\end{figure}

Fig.~\ref{fig:L_phase_equiv}(a) shows how a basic flux-aware inductor model could be constructed using a standard inductor and a current-controlled voltage source with a transimpedance of $g=L/\Phi_0$ Ohms. Similar considerations also apply for the mutual-inductance transformer, with current in the primary coil of the transformer generating flux across the secondary coil. Note that while these models correctly adjust the flux in accordance with the current flowing through the inductance, they do not draw current in the inductor in response to a voltage $V_{\Phi}$ being forced across the flux terminals. 

Fig.~\ref{fig:L_phase_equiv}(b) shows a more complete model following the ideas of Ref.~\onlinecite{shiri2023modeling}. The model includes a voltage-controlled current source with transadmittance of $G=-1$~S and an integrating capacitor $C_\Phi=2.068$~fF to set $V_{\Phi}\propto\int{V_Ldt}$ with the appropriate normalization, where $V_L$ is the voltage across the inductor terminals. The inductor current $I_L$ then relates to the flux via the $2\times2$ admittance matrix $Y$, $I_L=Y_{21}V_{\Phi}=V_\Phi\Phi_0/L$.

\begin{figure}[t]
    \includegraphics[width=\columnwidth]{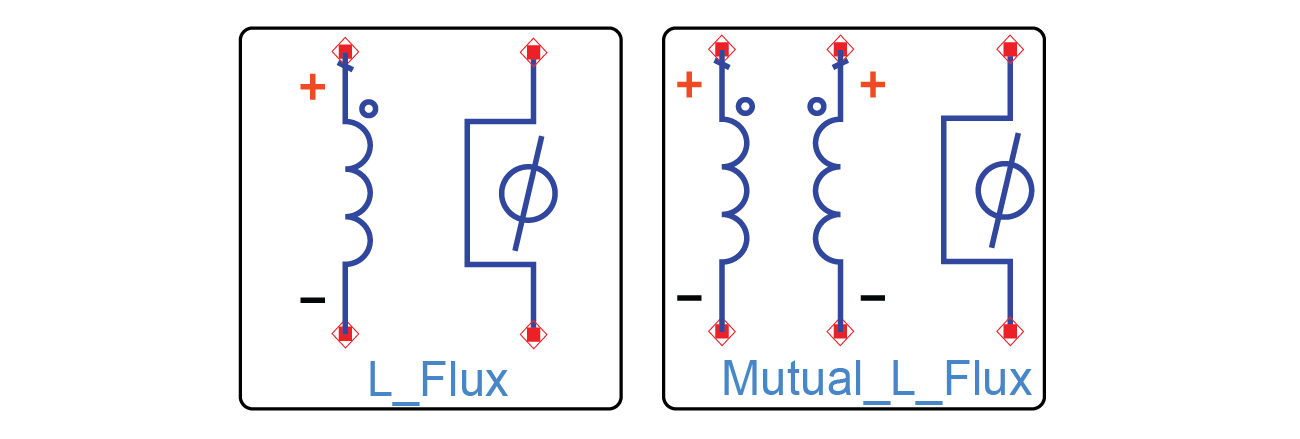}
    \caption{Flux-aware inductor and mutual-inductance models in ADS 2025.\label{fig:ind_models}}
\end{figure}

Keysight ADS 2025 introduces behavioral inductor and mutual inductance transformer models, depicted in Figure~\ref{fig:ind_models}. The (\texttt{L\_Flux}) model takes into account the flux across the inductance terminals, and has two modes of operation. The first mode, indicated by setting \texttt{Mode=current driven}, behaves similarly to Fig.~\ref{fig:L_phase_equiv}(a) and solves for voltages such that the voltage drop across the inductor flux nodes is $V_\phi = LI_L/\Phi_0$, and the voltage across the inductor terminals is the usual $V_L=L\frac{dI_L}{dt}$. The second mode, \texttt{Mode=voltage driven}, is more akin to Fig.~\ref{fig:L_phase_equiv}(b) and solves for currents such that the current in the inductor is $I_L=\Phi_0V_\phi/L$ and the ``current'' into the flux port is $I_\phi=\Phi_0\frac{dV_\phi}{dt}$. The two mode of operations for \texttt{L\_Flux} are mutually exclusive. The comprehensive model of Fig.~\ref{fig:L_phase_equiv}(b) can be easily constructed from standard library components but it doesn't permit mutually exclusive operational modes for the inductance.

The Mutual-Inductance with Flux (\texttt{Mutual\_L\_Flux}) component in Fig.~\ref{fig:ind_models} describes a mutual inductance $M$ between a primary inductance $L_1$ and a secondary inductance $L_2$ with the extra flux nodes properly accounting for the induced flux with a corresponding voltage drop such that
\begin{equation}
V_{\Phi\text{-ind}} = \frac{MI_1}{\Phi_0} + \frac{L_2I_2}{\Phi_0}.
\end{equation}
It is important to recognize here that there will be induced flux even at DC --  a unique feature for superconducting circuits. Meanwhile, currents and voltages in the arms of the mutual inductance device will follow
\begin{align}
V_1 &= L_1 \frac{dI_1}{dt} + M \frac{dI_2}{dt} \\
V_2 &= M \frac{dI_1}{dt} + L_2 \frac{dI_2}{dt}
\end{align}
where $V_1$, $I_1$, and $V_2$, $I_2$ are the voltage, current in $L_1$ and $L_2$, respectively.

\section{Flux Quantization in ADS}\label{sec:flux_quantization}
As already mentioned, flux quantization arises naturally in transient simulation starting from a quiescent initial condition that sets all source, currents, and voltages to zero. In other simulation modalities, such as S-parameter, DC, AC, and HB analyses, flux quantization must be enforced externally. In ADS, having access to the junction phases via $V_\phi$ and inductor fluxes via $V_{\Phi\text{-ind}}$, allows flux quantization to be expressed via an auxiliary circuit that constrains these fictitious voltages.

\subsection{Native flux quantization in transient simulation}
Since flux quantization is a consequence of electromagnetism, transient simulations with ADS native standard components and a realistic Josephson junction model should correctly capture the salient physics in superconducting electronic circuits. To test this, we use the \texttt{JJ3} model (see Figure~\ref{fig:jj_models}) to construct an rf-SQUID and measure its phase vs flux relation.
\begin{figure}[hb]
    \includegraphics[width=\columnwidth]{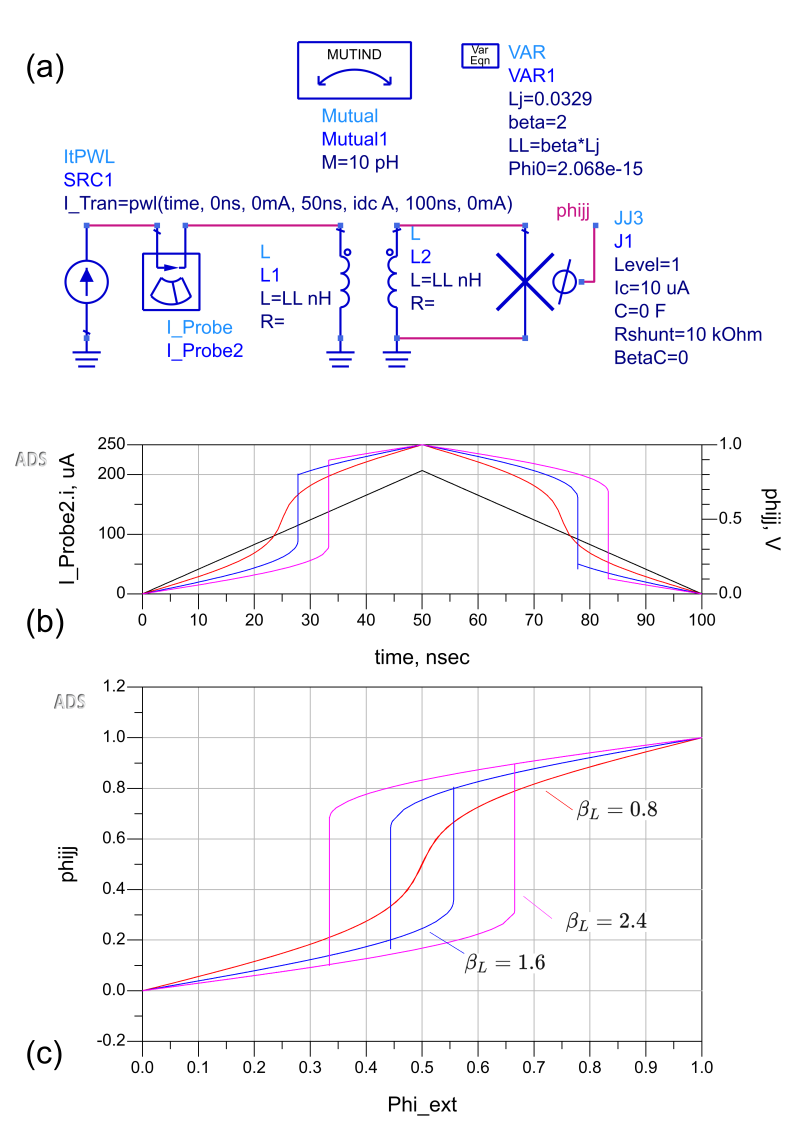}
    \caption{transient simulation of an rf-SQUID using the \texttt{JJ3} model. (a) circuit schematic, (b) current stimulus (left axis) and junction phase response (voltage at the junction phi node, right axis) as a function of simulated time, (c) junction phase vs applied flux for various $\beta_L$ values. \label{fig:rf_squid_trans}}
\end{figure}

Fig.~\ref{fig:rf_squid_trans}(a) shows the ADS circuit schematic used for this test. The rf-SQUID is the parallel connection of junction \texttt{J1}, having critical current $I_c=10\,\mu$A and inductor \texttt{L2}. The junction inductance is $L_{J0}=\hbar/2eI_c=32.9$~pH, and the linear inductance is $L_2=\beta_L\times L_{J0}$, where $\beta_L$ is a parameter swept in the simulation. For $\beta_L<1$ the rf-SQUID is mono-stable, but for $\beta_L>1$ the rf-SQUID can store flux and its phase vs flux relation becomes hysteretic. Flux is induced into the rf-SQUID via a transformer having inductor \texttt{L1} as its primary coil and \texttt{L2} as the secondary, with a mutual inductance $M=10$~pH represented by the component \texttt{Mutual1}, and driven by a piece-wise-linear current source \texttt{SRC1}.

Figure~\ref{fig:rf_squid_trans}(b) shows the current $I=$\texttt{I\_Probe2.i} in the primary coil (left y-axis) as a function of time, ramping from an initial condition of $I=0$ to a maximum of $I=$\texttt{idc}~A, corresponding to $1\,\Phi_0$ induced in the rf-SQUID loop: \texttt{idc}$=\Phi_0/M$, and then ramping back to zero. The figure also shows, on the right axis, the voltage $V_\phi$ on the node \texttt{phijj}, corresponding to the junction phase. The three traces correspond to three different values of $\beta_L=\left[0.8,\,1.6,\,2.4\right]$. Fig.~\ref{fig:rf_squid_trans}(c) shows the voltage $V_\phi$ at the \texttt{phijj} node plotted vs the externally applied flux bias, \texttt{Phi\_ext}$=MI/\Phi_0$ for the three values of $\beta_L$ as indicated in the figure. 

We observe the expected behavior of the rf-SQUID phase vs flux relation, which can be also calculated from
\begin{equation}\label{eq:rf_squid_current_phase}
    \phi+\frac{\beta_L}{2\pi}\sin{(2\pi\phi)}=\phi_\mathrm{ext},
\end{equation}
where $\phi_\mathrm{ext}$ is the external flux. For $\beta_L=0.8$ [red curve in Fig.~\ref{fig:rf_squid_trans}(c)], the phase vs flux relation is continuous and mono-stable but nonlinear, with the junction phase reaching $\pi/2$ ($V_\phi=0.25$) at \texttt{Phi\_ext}$\approx 0.38$. At this value of flux bias the junction inductance $L_J=L_{J0}/\cos{\left(2\pi\phi\right)}$ diverges. 

For higher values of $\beta_L$, the rf-SQUID phase vs flux relation becomes hysteretic  with the range of multistability increasing with $\beta_L$, and the junction phase $V_\phi$ (\texttt{phijj}) jumps discontinuously when its values reach $\phi_\mathrm{jump}$,
\begin{equation}\label{eq:rf_squid_jump}
    2\pi\phi_\mathrm{jump}=\cos^{-1}\left(-\beta_L^{-1}\right).
\end{equation}
For $\beta_L=1.6$ [blue in Fig.~\ref{fig:rf_squid_trans}(c)] these jumps occur at \texttt{Phi\_ext}$=0.444$ and $0.556$, and for $\beta_L=2.4$ (magenta in the figure) the jumps are at \texttt{Phi\_ext}$=0.334$ and $0.666$ as expected from Eqs.~(\ref{eq:rf_squid_jump}) and (\ref{eq:rf_squid_current_phase}).

\subsection{Flux quantization with an auxiliary circuit}
\begin{figure}[t]
    \centering
    \includegraphics[width=\columnwidth]{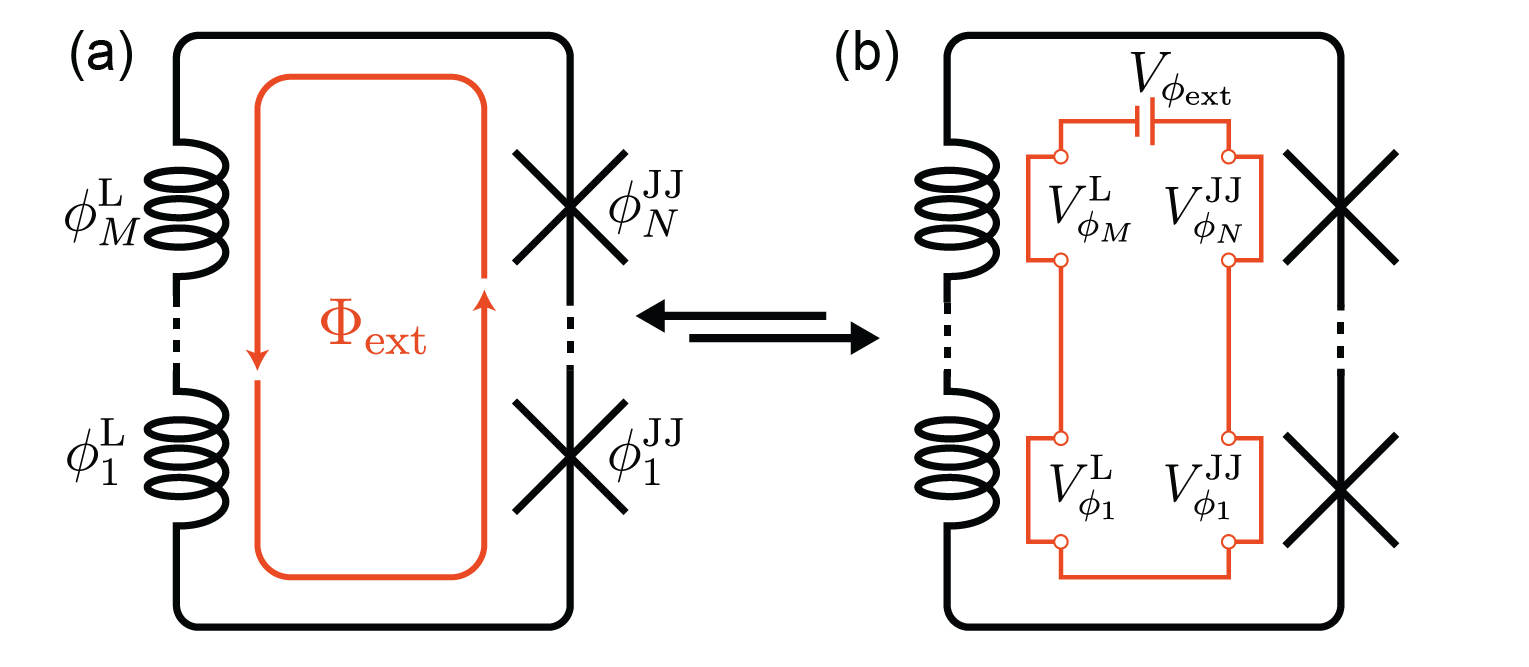}
    \caption{(a) Superconducting loop with $N$ Josephson junctions and $M$ inductors subjected to external flux $\Phi_\text{ext}$. (b) Corresponding circuit expressed in terms of voltage drops across each element and voltage bias via $V_{\phi_\text{ext}}$.}
    \label{fig:FluxQuantization}
\end{figure}

When a superconducting loop is subjected to an external magnetic field, the magnetic flux threading the loop is quantized in integer multiples of $\Phi_0$. For circuits with Josephson junctions and inductors, the flux quantization condition in the frequency domain boils down to a constraint on the sum of flux exhibited by the different components in the loop. Let's consider the superconducting loop shown in Figure \ref{fig:FluxQuantization} that contains an arbitrary number of Josephson junctions and inductors, $N$, and $M$, respectively. In the presence of external flux $\Phi_\text{ext}$, the sum of flux-induced phases around the loop should satisfy the relation 
\begin{equation}\label{eq:quantization}
\sum_{n=1}^{N} \phi^{\text{JJ}}_n + \sum_{m=1}^{M} \phi^{\text{L}}_m = \frac{\Phi_{\text{ext}}}{\Phi_0}+k,
\end{equation}
where $\phi^{\text{JJ}}_n$ and $\phi^{\text{L}}_m$ are the phase difference across Josephson junction $n$ and inductor $m$, respectively. The integer $k$ is the number of flux quanta enclosed in the loop: this number can be zero when $|\Phi_\mathrm{ext}|<\Phi_0/2$, but can change in circuits such as the dc-SQUID when the externally applied flux exceeds the above value; this change is associated with one of the junctions undergoing a $2\pi$ phase slip, generating a voltage pulse across the junction whose time integral is exactly $\Phi_0$.  

With the introduction of flux nodes for both the Josephson junctions and the inductors as was presented in Sections II and III, the flux condition can be translated to a voltage condition such that
\begin{equation}
\sum_{n=1}^{N} V_{\phi_n}^{\text{JJ}}+ \sum_{m=1}^{M} V_{\phi_m}^{\text{L}} = V_{\phi_{\text{ext}}},
\end{equation}
where $V_{\phi_n}^{\text{JJ}}$ and $V_{\phi_m}^{\text{L}}$ are the voltage drops across the phi terminals of the Josephson junction $n$ and inductor $m$, respectively.
Fortunately, this condition can be fulfilled with an auxiliary flux loop through the Kirchhoff voltage law as shown in Figure \ref{fig:FluxQuantization}. In principle, changing $V_{\phi_\text{ext}}$ is equivalent to changing the external flux which in turn will change the bias of the different junctions in the loop and their equivalent inductance effectively tuning the circuit. 

\begin{figure}[th]
    \includegraphics[width=\columnwidth]{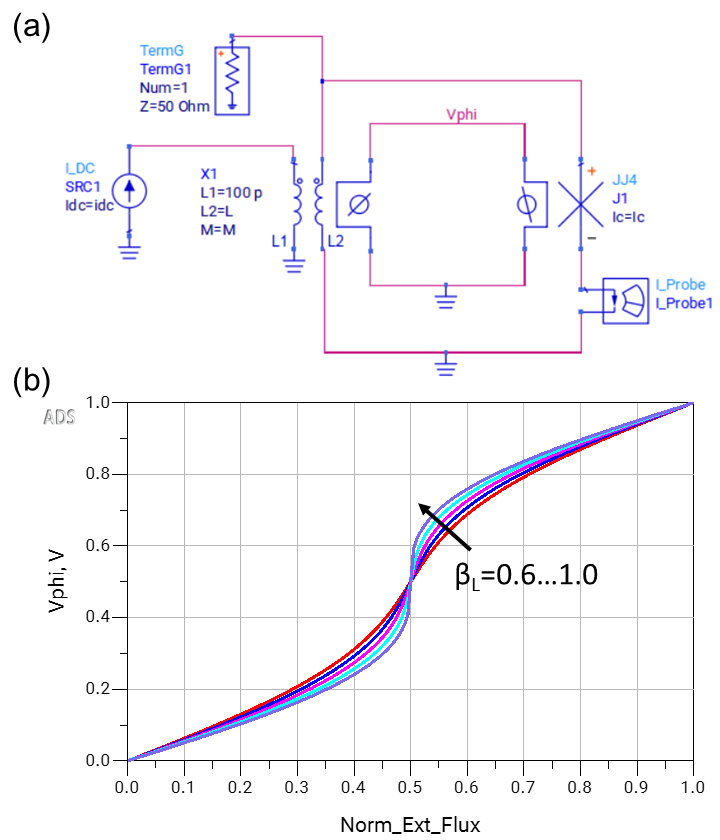}
    \caption{DC simulation of an rf-SQUID using the \texttt{JJ4} and \texttt{Mutual\_L\_Flux} models. (a) Circuit schematic. (b) Junction phi voltage vs. applied flux for various $\beta_L$ values.  \label{fig:rf_squid_dc}}
\end{figure}

\subsubsection{Single junction: rf-SQUID}
Figure~\ref{fig:rf_squid_dc} demonstrates how the flux quantization auxiliary circuit can be constructed for the simple case of an rf-SQUID. The schematic shown in Fig.~\ref{fig:rf_squid_dc}(a) is using the \texttt{JJ4} model and the secondary inductor of the \texttt{Mutual\_L\_Flux} model to form the rf-SQUID, and flux bias to the circuit is provided by the primary inductance of the \texttt{Mutual\_L\_Flux} component. In a DC simulation, the dc current source \texttt{SRC1} driving the flux bias is outputting $I=\mathtt{Norm\_Ext\_Flux}\times\Phi_0/M$, and \texttt{Norm\_Ext\_Flux}
is swept from 0 to 1.0. The flux quantization circuit is formed by connecting the phi port of the transformer in series with the phi port of the junction at the node \texttt{Vphi}. In the simulation, we sweep the value of $\beta_L$ from 0.6 to 1.0, and plot the voltage at the \texttt{Vphi} node (proportional to the junction phase) vs applied flux, \texttt{Norm\_Ext\_Flux}, in Fig.~\ref{fig:rf_squid_dc}(b). The resulting phase vs flux relation as seen in the figure agrees with the theoretical expected behavior of Eq.~(\ref{eq:rf_squid_current_phase}).

\begin{figure}[ht]
    \includegraphics[width=\columnwidth]{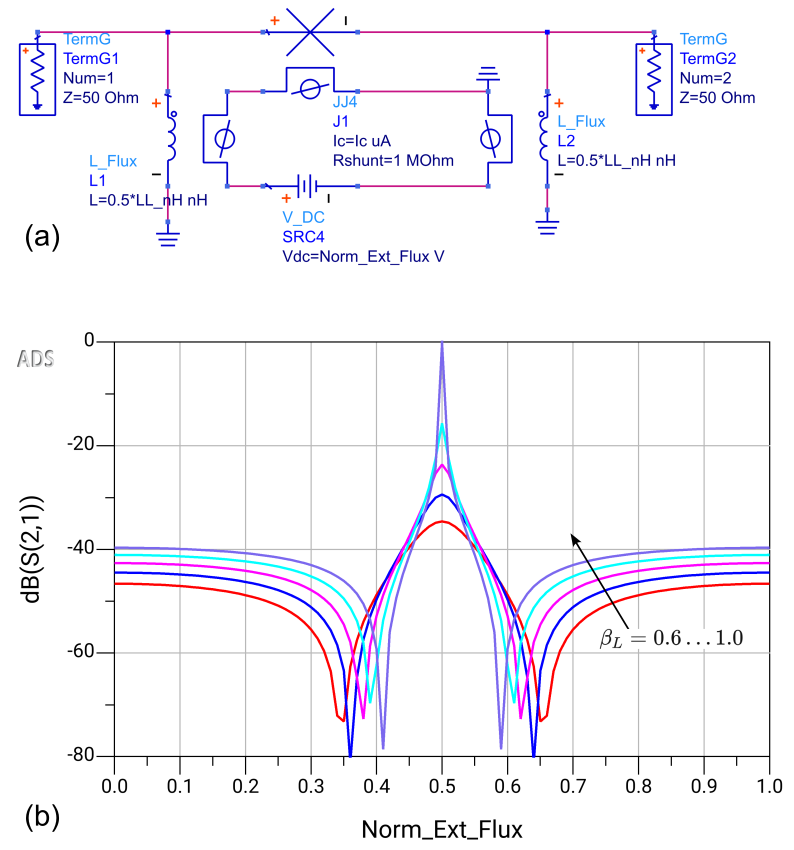}
    \caption{S-parameter simulation of an rf-SQUID coupler using a voltage source to directly drive the SQUID flux quantization circuit. (a) Circuit schematic. (b) $S_{21}$ vs applied flux for various $\beta_L$ values.\label{fig:rf_squid_coupler}}
\end{figure}

Alternatively, the flux threading the SQUID can be driven directly using a voltage source, as shown in Fig.~\ref{fig:rf_squid_coupler}(a). This circuit implements a tunable mutual coupler of the type used in Refs.~\onlinecite{chen2014qubit, naaman2016chip}: the coupling between the input and output ports can be nulled at a certain value of applied flux, for which the junction inductance diverges.

The schematic in Fig.~\ref{fig:rf_squid_coupler}(a) shows the linear inductor of the rf-SQUID split into two shunt branches bridged by a junction ($I_c=5\,\mu$A), forming a tunable inductive `pi' circuit. The transmission $S_{21}$ through the structure at 5~GHz is shown in Fig.~\ref{fig:rf_squid_coupler}(b) as a function of the voltage output by the source \texttt{SRC4} driving the flux quantization circuit (\texttt{Norm\_Ext\_Flux}) and for varying values of $\beta_L$ (which controls the value of the linear inductors in the circuit, as we have done in Fig.~\ref{fig:rf_squid_trans}). We see that indeed there is a null in the transmission at specific $\beta_L$-dependent values of the applied flux. The coupling near $\Phi_0/2$ ($\mathtt{Norm\_Ext\_Flux}=0.5$) is higher than that around integer normalize external flux, and becomes unity when $\beta_L=1$, all in accordance with the device theory. The hysteretic regime of $\beta_L>1$ is generally avoided in this type of application.

\subsubsection{Two junctions: dc-SQUID}\label{sec:dc_squid}
\begin{figure}[h]
    \includegraphics[width=\columnwidth]{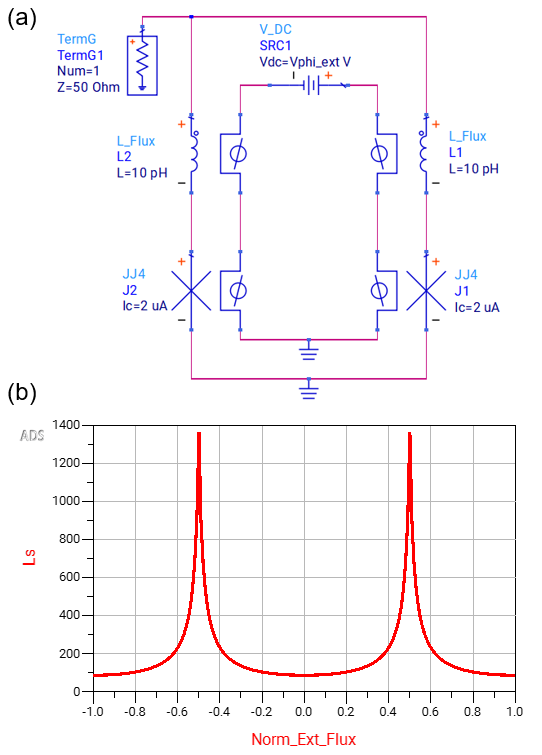}
    \caption{S-parameter simulation of a dc-SQUID using the \texttt{JJ4} and \texttt{L\_Flux} models. (a) Circuit schematic, (b) SQUID inductance in pH at 5~GHz, evaluated from the input impedance $L=\mathrm{Im}\{Z_\mathrm{in}\}/\omega$. \label{fig:dc_squid_study}}
\end{figure}

In a second simple example, we construct a dc-SQUID and extract its inductance from an S-parameter simulation. Fig.~\ref{fig:dc_squid_study}(a) shows the ADS schematic, where the SQUID is formed by the two \texttt{JJ4} and two \texttt{L\_Flux} blocks. The flux quantization circuit includes a dc voltage source, \texttt{SRC3}, to directly drive the external flux applied to the circuit.

We measure $S_{11}$ at $\omega/2\pi=5$~GHz, calculate the SQUID input impedance $Z_\mathrm{in}$ seen from port 1, and from it calculate the SQUID inductance $L=\mathrm{Im}\{Z_\mathrm{in}\}/\omega$. The simulation is repeated for different values of applied flux, from $-1.0\;\Phi_0$ to $1.0\;\Phi_0$, as represented by the quantity \texttt{Norm\_Ext\_Flux} in Fig.~\ref{fig:dc_squid_study}(b).

The dc-SQUID is a case where the number of flux quanta enclosed in the loop, $k$ in Eq.~(\ref{eq:quantization}), can change. To capture this in the simulation, we assign an expression to the variable \texttt{Vphi\_ext}, which sets the voltage of \texttt{SRC3} and captures the modularity of the enclosed flux, using an ADS conditional statement and the modulus \texttt{fmod()} function.
\begin{verbatim}
    Vphi_ext = 
      if Norm_Ext_Flux > 0 
        then fmod(0.5+Norm_Ext_Flux,1)-0.5 
        else fmod(-0.5+Norm_Ext_Flux,1)+0.5 
      endif
\end{verbatim}
Fig.~\ref{fig:dc_squid_study}(b) shows the simulated inductance $L_s$ in pH on the y-axis vs externally applied normalized flux, in agreement with the expected modulation curve of the SQUID. Because of the 20~pH linear inductance $L_\mathrm{loop}=L_1+L_2$ in the SQUID loop, the SQUID inductance does not truly diverge, and the range of modulation decreases for increasing $L_\mathrm{loop}/2L_J$ as expected and observed in simulations (not shown).

When $L_\mathrm{loop}>2L_J$ the dc-SQUID's inductance vs flux modulation curve can become multi-valued and hysteretic, and any asymmetry in the junction critical currents or in the branch linear inductances will further complicate the dynamics. To simulate such situations, we usually prefer to perform transient analysis to understand the system's stability regions; we can then perform small signal S-parameter or HB simulations in one of the stable regions.

\section{Compound circuits and arrays}
Many microwave Josephson circuits developed in recent years involve compound SQUIDs and SQUID arrays, in which flux quantization must be accounted for across several (and often nested) superconducting loops. These compound devices, such as the superconducting nonlinear asymmetric inductive element\cite{frattini20173, frattini2018optimizing, sivak2019kerr, miano2022frequency} (SNAIL), the asymmetrically threaded SQUID\cite{lescanne2020exponential} (ATS), and the snake inductor\cite{bell2012quantum, naaman2017josephson, white2022readout, kaufman2023josephson}, arise in an attempt by designers to engineer the nonlinearity of Josephson microwave circuits to gain new functionality or improve their power handling. Here we will examine the SNAIL and Snake devices in the context of parametric amplification, and show how to construct flux-quantization aware schematics in ADS to model their behavior.

\subsection{Kerr-free operation of a SNAIL amplifier}
\begin{figure}[th]
    \includegraphics[width=\columnwidth]{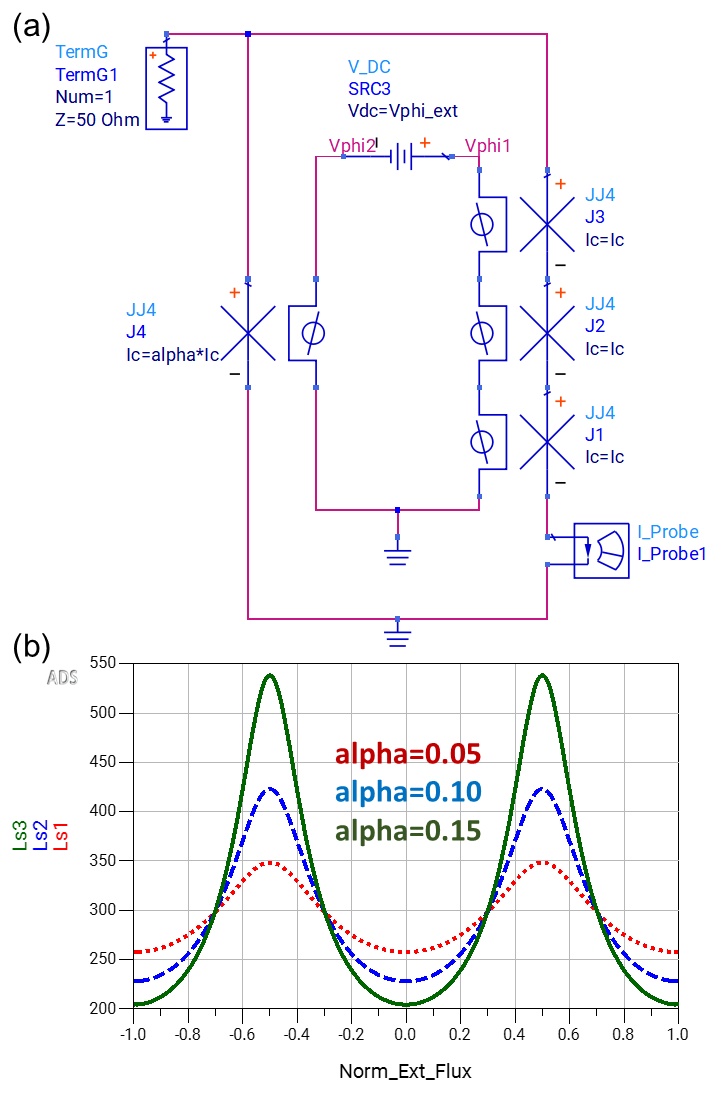}
    \caption{S-Parameter simulation of a SNAIL using the \texttt{JJ4} model. (a) Circuit schematic, (b) SNAIL inductance in pH at 5~GHz for $\alpha=0.05$, 0.1, and 0.15, evaluated from the input impedance $L=\mathrm{Im}\{Z_\mathrm{in}\}/\omega$. \label{fig:snail_dc_study}}
\end{figure}

The SNAIL is a device composed of a series array of three Josephson junctions with critical current $I_c$, in parallel with a smaller Josephson junction having a critical current of $\alpha I_c$, where $\alpha<1/3$, as shown in Fig.~\ref{fig:snail_dc_study}(a). The design parameter $\alpha$, together with a flux bias to the SNAIL loop, give the designer control of the third- and fourth-order nonlinear terms in the device's Hamiltonian. In particular, the device can be configured to feature a significant third-order nonlinearity that is useful for parametric amplification via a three-wave mixing process, while minimizing the fourth-order nonlinearity that is a contributor to saturation and harmonic distortion\cite{frattini20173, frattini2018optimizing, sivak2019kerr}.

Figure~\ref{fig:snail_dc_study}(a) shows an ADS schematic of a SNAIL device composed of four \texttt{JJ4} components. The flux quantization circuit is driven by a DC voltage source \texttt{SRC3}, setting the applied external flux to the SNAIL loop. Fig.~\ref{fig:snail_dc_study}(b) shows the simulated inductance of the SNAIL, computed from $S_{11}$ in an S-parameter analysis, as a function of flux bias and for various $\alpha$ values. The inductance varies periodically with flux, and modulates as expected with a range that increases with $\alpha$.

\begin{figure}[hbt]
    \includegraphics[width=\columnwidth]{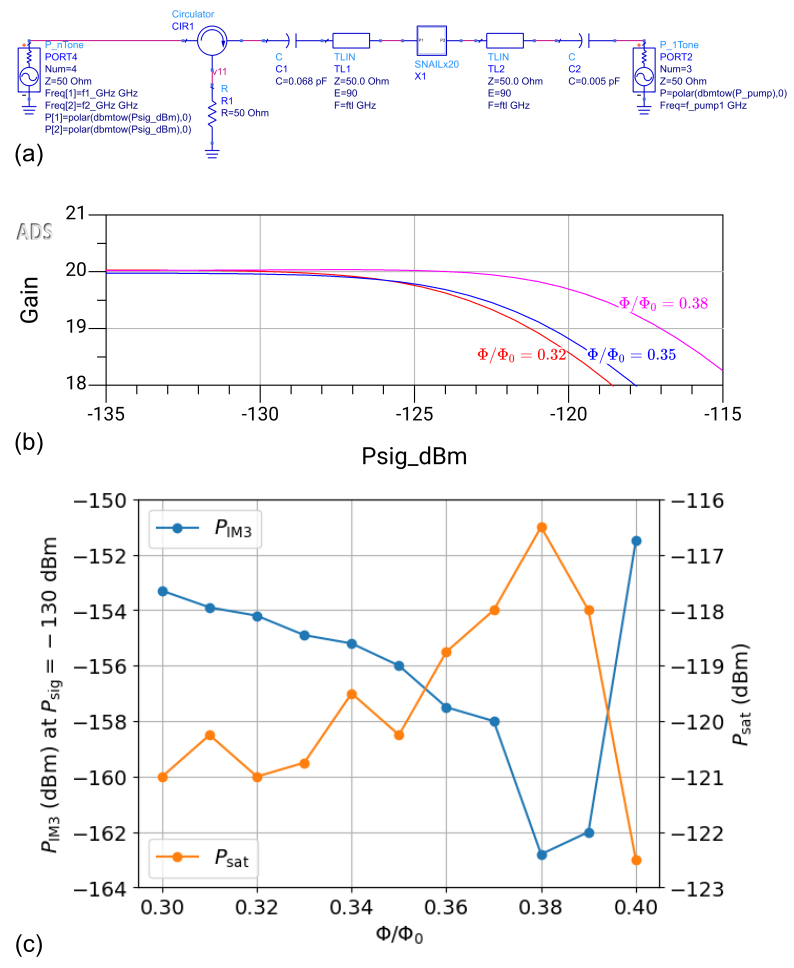}
    \caption{Harmonic balance simulation of a SNAIL amplifier based on `device C' in Ref.~\onlinecite{frattini2018optimizing}. (a) Circuit schematic, (b) amplifier gain vs input signal power in a 2-tone experiment for various flux bias points, (c) 2-tone saturation power (right axis, orange) and intermodulation product power (left axis, blue) as a function of flux. $P_\mathrm{IM3}$ was measured at input power of $-130$~dBm.\label{fig:snail_kerr}}
\end{figure}

In Fig.~\ref{fig:snail_kerr}, we implement a parametric amplifier based on a series array of 20 SNAIL elements (with all junction $R_\mathrm{shunt}=1\,\mathrm{M}\Omega$) embedded at the current antinode of a half-wave transmission line resonator\cite{frattini2018optimizing}. Fig.~\ref{fig:snail_kerr}(a) shows the circuit schematic; the SNAIL array with $I_c=8.5\,\mu$A and $\alpha=0.1$ is the sub-circuit \texttt{X1} in the figure, and transmission lines \texttt{TL1} and \texttt{TL2} form the resonator, whose frequency is tunable with flux between approximately 6 and 7.5 GHz. The signal is coupled into and out of the amplifier via capacitor \texttt{C1}, and the pump is fed via capacitor \texttt{C2}. Component values are shown in the figure, and are based on `device C' in Ref.~\onlinecite{frattini2018optimizing}.

To test the SNAIL model, we examine the so called `Kerr-free' point\cite{sivak2019kerr, frattini2018optimizing} of the SNAIL parametric amplifier by performing harmonic balance simulations with 2-tone drive, and measuring the amplifier's saturation power and 3rd order intermodulation products. At the flux bias corresponding to the Kerr-free point, the fourth-order nonlinearity of the amplifier is minimized giving rise to higher saturation power and lower harmonic distortion. At each simulated flux bias, we adjust the pump frequency to be twice the resonant frequency of the SNAIL-loaded resonator, and adjust the pump power to obtain 20 dB of gain. The two signal tones are separated by 100~kHz and are centered at a 500~kHz detuning from the center frequency of the resonator.

Fig.~\ref{fig:snail_kerr}(b) shows the simulated gain vs input signal power (per tone) for 3 different flux biases. We observe the familiar gain compression as the signal power is increased, with a flux-dependent $P_\mathrm{1dB}$ compression point. Near $\Phi/\Phi_0=0.38$, the amplifier's saturation power reaches a maximum of $-117$~dBm per tone, or $P_\mathrm{1dB}\sim-114$~dBm of total power, in reasonable agreement with the experiments in Ref.~\onlinecite{frattini2018optimizing}.

Fig.~\ref{fig:snail_kerr}(c) shows $P_\mathrm{1dB}$ (orange, right axis), and the power in the 3rd order intermodulation product $P_\mathrm{IM3}$ (blue, left axis). $P_\mathrm{IM3}$ was recorded for input signal power of $-130$~dBm, well below saturation. The figure clearly shows a maximum in the saturation power and a coincident minimum in the 3rd order intermodulation product power near $\Phi/\Phi_0=0.38$, which is identified as the Kerr-free operating point in Ref.~\onlinecite{frattini2018optimizing} for a device with similar parameters.

\subsection{Snake rf-SQUID array and the Lumped-Element Snake Amplifier}
\begin{figure}[t]
    \includegraphics[width=\columnwidth]{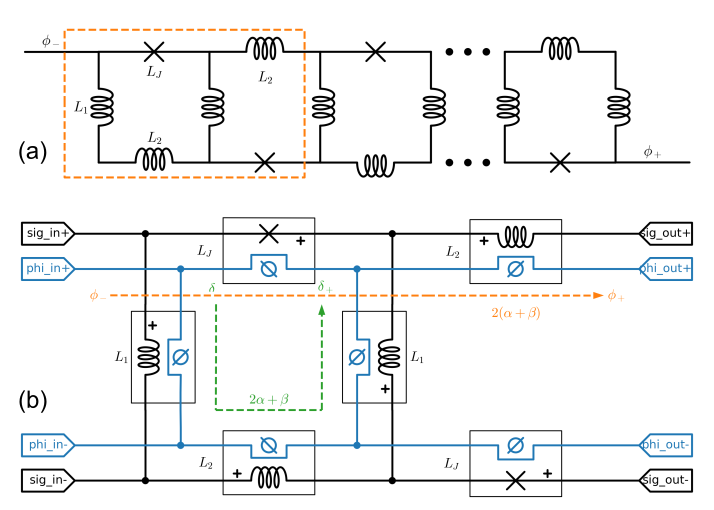}
    \caption{(a) Schematic of a snake array. (b) Schematic of a 2-junction stage of the Snake element\cite{white2022readout}, as indicated by the dashed box in panel (a). The signal path is indicated in black, while the flux-quantization circuit is highlighted in blue. Dashed green arrow indicates the phase advance around the rf-SQUID loop, and dashed orange arrow indicates phase advance across the array stage. Inductors $L_1$ and $L_2$ are using the model of Fig.~\ref{fig:L_phase_equiv}(b). \label{fig:snake_schematic}}
\end{figure}

The `snake' is a nonlinear inductance made with an array of interleaving rf-SQUIDs, as shown in Fig.~\ref{fig:snake_schematic}(a). It is composed of a meandering spine of alternating linear inductors, $L_1$ and $L_2$, and each of the meanders is bridged by a Josephson junction with inductance $L_J$, such that $L_J>4L_1+L_2$. This structure is based on an array proposed in Ref.~\onlinecite{bell2012quantum}, and was used (with linear inductances replacing Josephson junctions in the spine) in Ref.~\onlinecite{naaman2017josephson} as a high saturation power Josephson nonlinear element.

Flux is quantized in each of the rf-SQUID loops in the array. If $\alpha$ is the normalized flux due to current in inductance $L_1$, $\beta$ is the flux across $L_2$, and $\delta$ is the flux across the junction, then in each loop containing two $L_1$ inductors, an $L_2$ inductor, and a junction, we have $2\alpha+\beta=\delta$. The array is driven by imposing a phase difference $\phi=\phi_+-\phi_-$ across the $N$-stage array [see Fig.~\ref{fig:snake_schematic}(a)], such that $\phi=N\left(\alpha+\beta\right)$.

In ADS, we can define a sub-circuit to represent a 2-junction stage of the snake inductor, as highlighted with a dashed box in Fig.~\ref{fig:snake_schematic}(a). The schematic of that sub-circuit is shown in Fig.~\ref{fig:snake_schematic}(b), and uses the \texttt{JJ4} component (with $R_\mathrm{shunt}=1\,\mathrm{M}\Omega$) and the \texttt{L\_Flux} model of Fig.~\ref{fig:L_phase_equiv}(b). In the figure, the signal path is indicated with black wires, and the auxiliary flux quantization circuit is indicated with blue wires. In the rf-SQUID loop (green dashed arrow), the voltages $V_{\phi\text{-ind}}$ associated with the flux across the inductors, are summed and applied across the junction $V_\phi$ terminals. The flux across the 2-junction stage is propagated from the input to the output phi terminals of the block, advancing as indicated by the dashed orange arrow. The snake array can then be constructed by connecting a number of these sub-circuits in series, and terminating the array's last stage with an additional \texttt{L\_Flux} component.

\begin{figure*}[ht]
    \includegraphics[width=\textwidth]{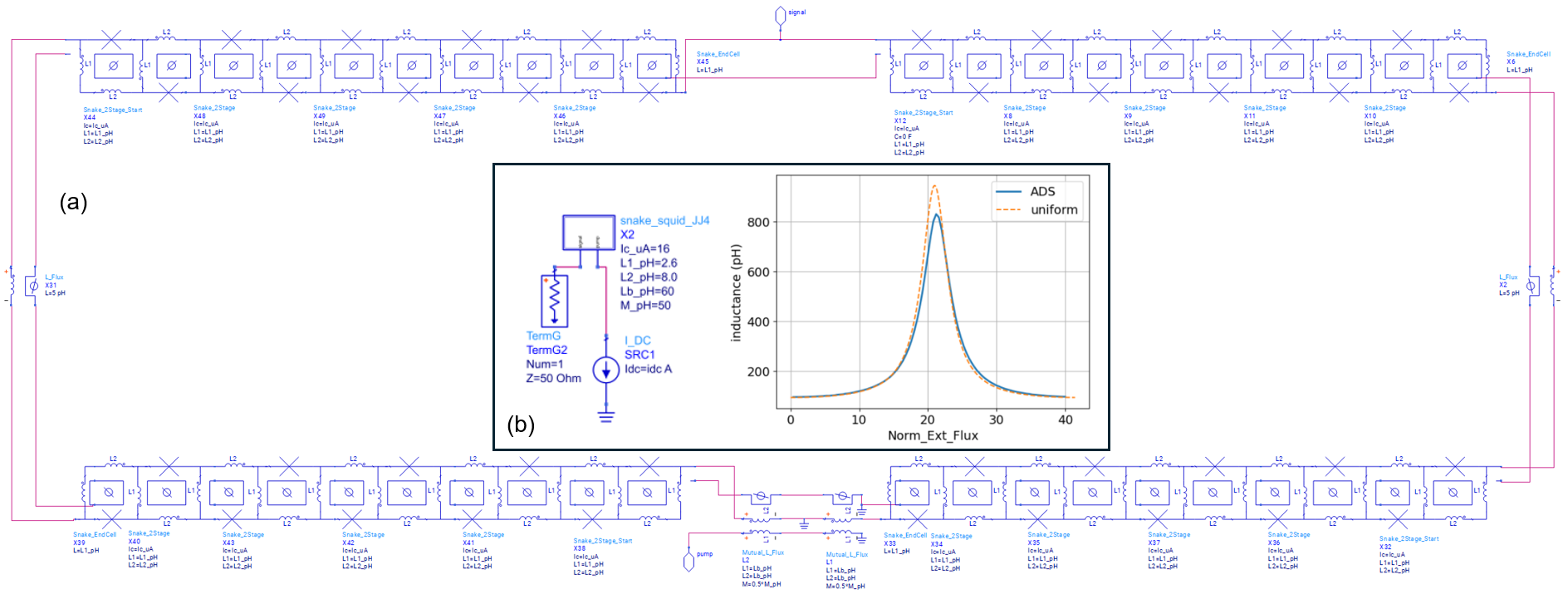}
    \caption{(a) Schematic of a snake array with total of 40 junctions. (b) Inductance of the snake array vs applied normalized flux, comparing ADS simulation to a closed-form expression assuming a uniform array, Eq.~(\ref{eq:snake_ind_thy}). Inset: ADS schematic used in simulation, component \texttt{X2} is the snake array shown in panel (a). \label{fig:snake_inductance}}
\end{figure*}

Figure~\ref{fig:snake_inductance}(a) shows a nonlinear inductance element built from 20 such sub-circuits, a total of 40 junctions, arranged in two parallel arrays of $N=20$ rf-SQUID stages each. We call the common mode of this array the `signal' mode. The differential `circulating' mode of the circuit includes a superconducting transformer (two \texttt{Mutual\_L\_Flux} components in series in the figure), which is driving a flux bias to the circuit via a mutual inductance $M$. The particular snake array shown in Fig.~\ref{fig:snake_inductance}(a) is modeled according to that used in Refs.~\onlinecite{white2022readout, kaufman2023josephson}, with $L_1=2.6$~pH, $L_2=8$~pH, and junction $I_c=16\,\mu$A.

We perform S-parameter simulations at a fixed frequency of 5~GHz, and with a swept applied external flux. Similar to what we have done in Sec.~\ref{sec:dc_squid}, we extract the array's inductance as seen from the signal port from the simulated $S_{11}$, and plot this inductance vs flux bias in Fig.~\ref{fig:snake_inductance}(b), solid blue trace. The simulated schematics is shown in panel (b) inset, with sub-component \texttt{X2} representing the snake array of panel (a), and current source \texttt{SRC1} providing a dc flux bias across the entire 40-squid array by outputting $I_\mathtt{idc}=\mathtt{Norm\_Ext\_Flux}\times\Phi_0/M$.

Fig.~\ref{fig:snake_inductance}(b) additionally shows, in dashed orange line, the calculated inductance
\begin{equation}\label{eq:snake_ind_thy}
    L_\mathrm{snake}=\frac{N}{2}\frac{\left(L_1+L_2\right)L_J+L_1L_2\cos\delta_0}{L_J+\left(4L_1+L_2\right)\cos\delta_0},
\end{equation}
where $\delta_0$ is the equilibrium phase across each of the junctions\cite{naaman2017josephson, white2022readout}, and where the flux axis is scaled to account for the bias transformer inductance according to Eq.~(S12) in Ref.~\onlinecite{white2022readout} appendix. Eq.~(\ref{eq:snake_ind_thy}) assumes the snake array is uniform and perfectly periodic, and therefore represents an approximation neglecting any nonuniformity of the currents arising from the finite extent of the array. As we can see from Fig.~\ref{fig:snake_inductance}(b), the uniform-array approximation is in reasonable agreement with the simulation results. This example shows that while a closed-form approximation is useful in designing such devices, only circuit simulations can fully capture subtleties that can be important to the exact flux dependence of the inductance, especially in the context of parametric amplification and higher-order nonlinearities that contribute to intermodulation distortion and saturation. 

\begin{figure}[hbt]
    \includegraphics[width=\columnwidth]{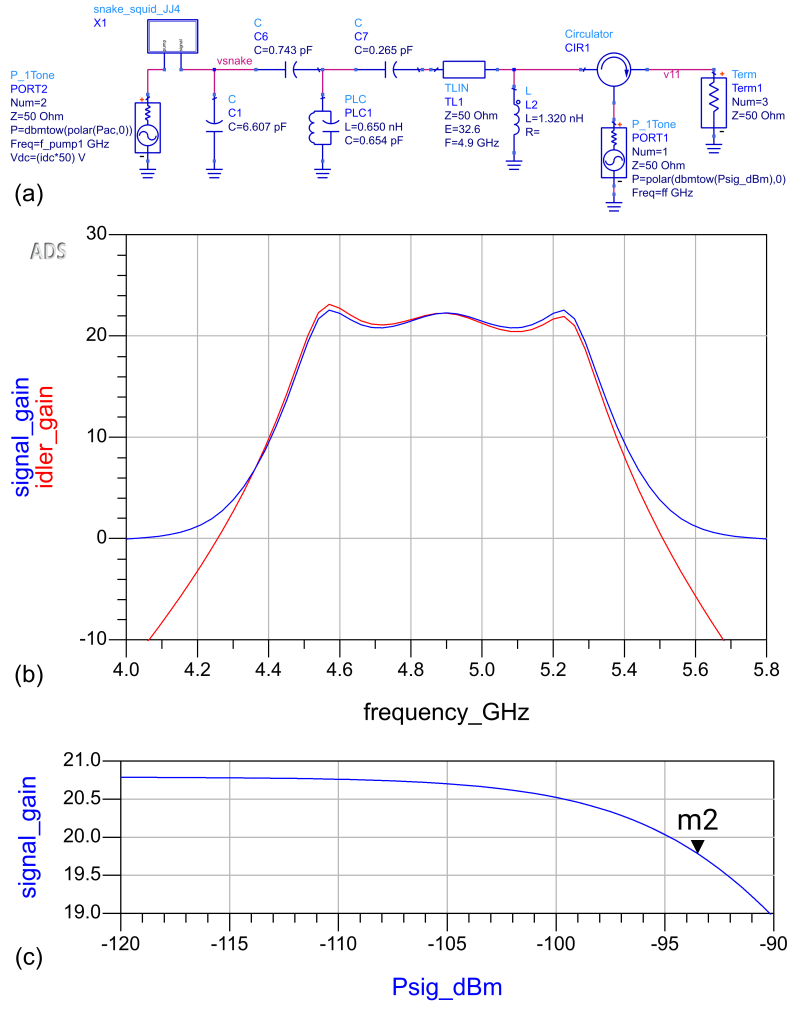}
    \caption{(a) Schematic of a LESA wide-band parametric amplifier. (b) Signal and idler gain vs frequency from HB simulation, (c) Signal gain vs input signal power. \label{fig:LESA_simulation}}
\end{figure}

Figure~\ref{fig:LESA_simulation}(a) shows an ADS schematic of the lumped-element snake amplifier (LESA) demonstrated in Ref.~\onlinecite{kaufman2023josephson}. This is a broad-band matched\cite{naaman2022synthesis} Josephson parametric amplifier based on a 3-pole Chebyshev matching network, and designed to have high saturation power by using the snake inductor as its nonlinear element. The schematic uses the same snake component shown in Fig.~\ref{fig:snake_inductance}(a) and (c), pumped and biased by source \texttt{PORT2} in the figure, outputting $-35.5$~dBm at $f_p=9.8$~GHz (modulation amplitude of $1.81\,\Phi_0$), and with a DC level set to bias the snake with a flux of $13.5\,\Phi_0$. The input signal, with input power of \texttt{Psig\_dBm}, is provided by \texttt{PORT1} through an ideal circulator. The power delivered to the load \texttt{Term1} is measured in a harmonic balance simulation at both the signal frequency, $f_s$, and the idler frequency, $f_i=f_p-f_s$. All component values are given in Fig.~\ref{fig:LESA_simulation}(a).

Fig.~\ref{fig:LESA_simulation}(b) shows the signal gain in dB (blue), extracted from the HB simulation from the voltage at the node \texttt{v11}, using the ADS \texttt{mix()} function,
\begin{verbatim}
    signal_gain=dBm(mix(v11, {0, 1}))-Psig_dBm
\end{verbatim}
where the indexes \texttt{\{0, 1\}} indicate a frequency that is the sum of the zeroth harmonic of the pump and the first (fundamental) harmonic of the signal. The idler gain (red) can be extracted similarly by using indexes \texttt{\{1, -1\}}, selecting a frequency that is the difference between the pump and signal fundamentals frequencies. We see in the figure that the small-signal ($\mathtt{Psig\_dBm}=-120$~dBm) gain of the LESA amplifier is just over 20~dB, with approximately 1~dB gain ripple, and a bandwidth exceeding 600~MHz.

Fig.~\ref{fig:LESA_simulation}(c) shows a HB simulation in which the signal frequency was fixed at 4.7~GHz and the signal power was swept. We see the familiar gain compression phenomenon, with a 1-dB input compression power of $P_\mathrm{1dB}=-93.5$~dBm (indicated in the figure by marker \texttt{m2}, in agreement with the experimentally measured LESA devices in Ref.~\onlinecite{kaufman2023josephson}.

\section{Working with flux models}
Representing the flux in a circuit as a fictitious voltage requires some care when working with flux-aware models. In inductance models such as the one in Fig.~\ref{fig:L_phase_equiv}(a), where the flux port is driven by what is essentially an ideal voltage source with zero internal resistance, a case where such inductors are placed in parallel could result in either voltage source attempting to bias an effective short circuit. The simulator will attempt a solution with unusually large currents flowing through the flux loop and will often fail, or may attempt to `fix' the problem by adding a small series resistance, which will then change the applied flux in the circuit. Such topolgies should be avoided, or otherwise the model of Fig.~\ref{fig:L_phase_equiv}(b) should be used instead. Additionally, the flux port and the signal port of the inductor and junction components are not independent, and nonsensical results could be obtained if we build a circuit that attempts, for example, to both current-bias and phase-bias the component simultaneously. The \texttt{L\_Flux} models shipped with ADS can help designers guard against these unintended behaviors by exclusively limiting the model's operation to either a voltage (flux) bias mode or a current bias mode.

The use of a voltage source to implement a flux bias in a circuit can also tempt a designer to simulate circuits that cannot be physically implemented. For example, in simulation it would be perfectly legal to directly phase bias a junction that is not embedded in a superconducting loop, essentially using the $V_\phi$ source as a `flux battery'. However, no such device exists in the physical world. Direct flux bias is only valid within a context of a superconducting loop, as we demonstrated in Figs.~\ref{fig:rf_squid_coupler}-\ref{fig:snail_dc_study}. Biasing a physical design via a mutual transformer as in Figs.~\ref{fig:rf_squid_dc} and~\ref{fig:snake_schematic} is in many cases preferred, as it anchors the simulation in physical reality. On the other hand, direct access to a supreconducting circuit's flux nodes offers some new opportunities in characterizing devices in simulation\textemdash enabling, for example, a direct probe of a device's nonlinearity, current-phase relation,  or inductance flux-tuning curve in isolation.  

\section{Conclusions}
We have introduced Josephson junction and inductance models in Keysight ADS, which explicitly account for the node fluxes at the devices' terminals, and expose them on the device schematic symbols as auxiliary port. Using these models, it is now possible to enforce flux quantization conditions in ADS simulations of superconducting devices\textemdash this is a critical feature whose absence thus far has limited the usability of modern EDA tools in microwave superconducting circuit design. We have shown a few examples of how such circuits can be constructed, going from simple devices like the rf- and dc-SQUID to multi-junction compound devices and arrays. We validated the models and their usage in DC, S-parameter, and harmonic balance simulations, against theoretical results as well as published experimental data. In particular, we demonstrated an important use case in the simulation of Josephson parametric amplifiers, and reproduced in simulation some key experimental results from SNAIL and LESA devices. We are excited by the opportunities that these methods offer to modernize the simulations, and accelerate the development of microwave superconducting electronic devices and circuits.

%

\end{document}